\documentclass[conference]{IEEEtran}
\IEEEoverridecommandlockouts
\usepackage{cite}
\usepackage{amsmath,amssymb,amsfonts}
\usepackage{algorithmic}
\usepackage{graphicx}
\usepackage{textcomp}
\usepackage{xcolor}
\def\BibTeX{{\rm B\kern-.05em{\sc i\kern-.025em b}\kern-.08em
    T\kern-.1667em\lower.7ex\hbox{E}\kern-.125emX}}
\begin{document}

\title{A GPT-based Code Review System for Programming Language Learning}

\author{\IEEEauthorblockN{Lee Dong-Kyu}
\IEEEauthorblockA{\textit{Department of Electrical Engineering and Computer Science,
}\\{University of Hanyang}
\\{Seoul, Republic of Korea}
}}

\maketitle

\begin{abstract}
The increasing demand for programming language education and growing class sizes require immediate and personalized feedback. However, traditional code review methods have limitations in providing this level of feedback. As the capabilities of Large Language Models (LLMs) like GPT for generating accurate solutions and timely code reviews are verified, this research proposes a system that employs GPT-4 to offer learner-friendly code reviews and minimize the risk of AI-assist cheating.

To provide learner-friendly code reviews, a dataset was collected from an online judge system, and this dataset was utilized to develop and enhance the system’s prompts. In addition, to minimize AI-assist cheating, 
 the system flow was designed to provide code reviews only for code submitted by a learner, and a feature that highlights code lines to fix was added. After the initial system was deployed on the web, software education experts conducted usability test. Based on the results, improvement strategies were developed to improve code review and code correctness check module, thereby enhancing the system. 

The improved system underwent evaluation by software education experts based on four criteria: strict code correctness checks, response time, lower API call costs, and the quality of code reviews. The results demonstrated a performance to accurately identify error types, shorten response times, lower API call costs, and maintain high-quality code reviews without major issues. Feedback from participants affirmed the tool’s suitability for teaching programming to primary and secondary school students. Given these benefits, the system is anticipated to be a efficient learning tool in programming language learning for educational settings.
\end{abstract}

\begin{IEEEkeywords}
Large Language Models (LLMs), GPT-4, Programming Language Education, Learner-Friendly Code Reviews
\end{IEEEkeywords}

\section{Introduction}
\subsection{Background}
Digital Transformation across all sectors of society is currently having a positive impact on various socio-economic areas. At this point of Digital Transformation, a global trend is forming that emphasizes the importance of education to equip everyone with “Digital Competencies” so they can participate in and benefit evenly from the digital world [1]. According to the Organization for Economic Cooperation and Development Skill Outlook 2019 report [2], Digital competencies will be closely related to all jobs in the future, so that there is a need to cultivate comprehensive digital skills throughout life.

In response to this global trend, Educational institutions in Korea have also enhanced education in order to foster digital and artificial intelligence(AI) literacy through National Curriculum of 2022 revision from Ministry of Education. Particularly in primary and secondary schools, Computer and Information education is being expanded by increasing either Computer and Information class or discretionary time [3]. While Computer and Information education has become mandatory in primary and secondary schools, Educational institutions can allow the introduction of various coding courses tailored to students’ career paths and aptitudes. Primary schools can now organize and operate Practical Science for previously 17 hours, but now over 34 hours. Secondary schools can organize and operate Computer and Information subject for previously 34 hours, but now over 68 hours. In addition, most core subjects such as Korean language, Science, Social Studies, and Art are increasingly related with contents that promote Digital competencies. 

\subsection{Challenge}
To strengthen digital competences in primary and secondary schools, learners have many opportunities to access online and offline programming education environments. Accordingly, the demand for programming language education is increasing, and the size of classes that must be taught in one class is also increasing. However, due to the nature of programming language education, various solutions are possible during learning. Since each learner has a different implementation method and a variety of code written, providing personalized code reviews can be an effective approach.

However, traditional code review methods have limitations in providing immediate and personalized feedback [4], [5]. Automated test case method is provided through online code correctness check systems, but it only point out issues such as compilation errors or inconsistencies in input/output examples, which can be difficult for beginners to understand.

Peer review methos conducted through online forums requires significant time and effort to provide appropriate solutions and can be challenging due to varying levels of learners [6].

The tutor review is a method in which learners mostly rely on the instructor to receive code reviews. Although it is an effective way to provide personalized and customized code reviews, it is very difficult for instructors to provide individual support due to the nature of the offline environment where many students must be managed. 

\subsection{GPT in Programming Language Learning}
The recent emergence of GPT, a Generative Pretrained Transformer model among large language models (LLM), provides an effective solution that overcomes the limitations of traditional code review methods [6]. In addition, Codex, a language model that analyzes the natural language description requested by the user as a prompt and converts it into code, is installed in GPT and is used as a powerful programming support tool. These powerful features of GPT show that GPT has good performance not only in text generation but also in code analysis and generation, becoming an appropriate approach for immediately performing code reviews and providing personalized comments in a programming language learning support system.

However, as the performance of LLM is upgraded and its use spreads in industrial section, many concerns are raised about using LLM in an educational environment due to the characteristic of LLM that provides immediate answers to any type of problem. In particular, there are concerns about “AI-assist cheating” side effects that may occur when students directly use ChatGPT, a service that allows easy use of GPT in introductory programming courses. Therefore, for effective real-time support solutions for programming language courses, it is necessary to construct a system that can perform immediate and effective code reviews while minimizing AI-assist cheating.

Therefore, to develop pedagogically effective, real-time support solutions for programming language courses, it is crucial to create systems that can provide immediate and effective code reviews while minimizing the risks associated with AI-assist cheating. This approach will ensure that learners not only receive the support they need but also engage in their learning processes authentically and responsibly. 

\subsection{Research Purpose}
In this paper, I present a code review system using GPT-4 that supports programming language learning for primary and secondary school students. This system aims to minimize AI-assist cheating and provide learner-friendly feedback appropriate for the target age. To minimize AI-assist cheating, code review comments are provided only for code submitted by learners, and code solutions are not provided directly in comments. 

To provide learner-friendly feedback, prompts were designed considering the precision, usefulness, specificity, supportive tone, and learning effect of code review comments. The system provides students with a bank of programming exercises. When a student selects a problem to solve and submits code, the system shows the correctness of the code and a code review comment on the students’ submitted code. It is an integrated programming language learning support environment.

This system was developed considering deployment in an introductory Python Programming course for elementary, middle, and high school students. A total of two system evaluations were conducted to evaluate the system, and current instructors with more than 2 years of programming education experience were selected as participants. 
Based on the results of the first system evaluation, the system advancement strategy for this study was summarized and it was improved.

In order to effectively test the performance of the improved system’s code correctness check and code review comment, it was evaluated according to the following research queries.

\begin{itemize}
    \item \textbf{RQ1 – Strict Code Correctness Check in Improved System}: Are the results of the GPT-based code correctness check more strict than traditional online judge system?
    \item \textbf{RQ2 – Response Time Reduced in Code Reviews}: Does the improved system reduce the response time for code review comments compared to the existing system?
    \item \textbf{RQ3 – Cost Reduction via API Call Optimization}: Does the improved system reduce API call costs compared to existing system?
    \item \textbf{RQ4 – Quality of Code Reviews}: Despite reduced response times and API call costs, does the system maintain the quality of code reviews?
\end{itemize}

\section{Related Work}
\subsection{LLMs in Educational Feedback}
As LLMs become widely used in practical applications, educational experts are exploring the potential of using GPT to generate educational feedback on student assignments. Wei Dai et al. [7] collected business scenario-related data science project proposals from students at an Australian university. The proposals were required to include a project description and business model. Instructors evaluated the submissions based on five criteria: objectives, topic relevance, business benefit, novelty, and clarity, providing text feedback subsequently. After anonymizing the students’ personal information, feedback related to 103 students proposals was compiled.

To generate text feedback on these proposals using ChatGPT, the prompt was designed with properties: project’s objectives, relevance to the data science topic, business benefit, novelty/creativity, overall clarity. Each student’s proposal text was added after the prompt and submitted to ChatGPT, which generated the feedback. Interestingly, it was noted that the instructors’ feedback was on average 109 characters less than that provided by ChatGPT. While a more word count does not necessarily equate to more effective feedback, three experts rated each piece of feedback for readability and consistency on a scale from 0 to 4, finding that ChatGPT’s feedback was more detailed and readable than that of the instructors. 

GPT could potentially replace some of the instructor’s workload while providing useful text feedback to students in educational settings.

\subsection{Code Review Automation on Basic Programming training}
Learning tools that provide automated code feedback, such as the online judge service, deal with code submitted by students in a test case manner. This method runs the students’ programs against test cases and then provides feedback on failed cases. The error message is in English and consists of terms related to compilation errors, making it difficult for beginners to understand the cause of the issue.

To provide automatic feedback to novice learners, Rishabh Singh [8] proposed a new method in which learning tools require error information from the instructor, including solutions and potential corrections for errors students may make to the method. Based on this information, the system can automatically identify the minimal corrections to a student’s incorrect code and provide feedback on what went wrong, along with a quantifiable measure of how wrong a submitted code is.
	
The feedback provided should consist of the following four pieces of information: the location of the error as a line number, the problematic expression on that line, the sub-expression that needs to be modified, and the newly modified value of the sub-expression. To provide this level of feedback, an automated feedback system requires correction rules from the instructor that indicate not only the problems students need to solve and solution, but also ways to correct the types of error students may make. The learning tool then applies these rules to the student program, searches for candidate programs that match the correct answer code and requires minimal modification, and provides feedback.

The composition of the feedback provided to novice learners and the information required from the instructor to generate the feedback are factors that must be sufficiently addressed when designing prompts for an LLM-based code review system.

\subsection{Prompt Template for Code Review Automation}
Compared to pre-trained language models (PLMs), LLMs in code review automation provide a more efficient and general-purpose approach to complex task such as code review automation. Because LLM already internalizes various domain knowledge, it requires fewer resources for domain-specific prior training and can be flexibly applied to a variety of tasks, contributing greatly to the development of code review automation.

To automate code reviews, J. Lu et al. [9] proposed a pipeline as a cycle of the code review process that predicts the need for reviews, generates review comments, and sequentially performs code improvement tasks.

The three core tasks essential to the code review process are:
\begin{itemize}
    \item \textbf{Review Necessity Prediction (RNP)}: A task that responds with a binary label (yes/no) whether a diff hunk requires a review
    \item \textbf{Review Comment Generation (RCG)}: Generates a comment for a given diff hunk. Two perspectives are presented here: a line-level perspective, which focuses on the content of individual lines of code (using Crer dataset), and a method-level perspective, which provides a holistic view of the code context (using Tufano dataset)
    \item \textbf{Code Refinement (CR)}: Code refinement involves minor adjustments or rearrangements to existing code to improve the quality of the code. Due to these minor modifications, the input code and output code often show strong similarities
\end{itemize}

The template containing the prompt request and response utilized Stanford Alpaca’s template and following the format: instructions, input(optional), output by clearly distinguishing and modularizing each step is a systematic approach. 

In terms of systematically managing prompt templates in this way, the maintenance task become efficient and effective by fixing the response results of LLM and updating the prompts in specific areas.

\begin{figure} [h]
    \centering
    \includegraphics[width=0.75\linewidth]{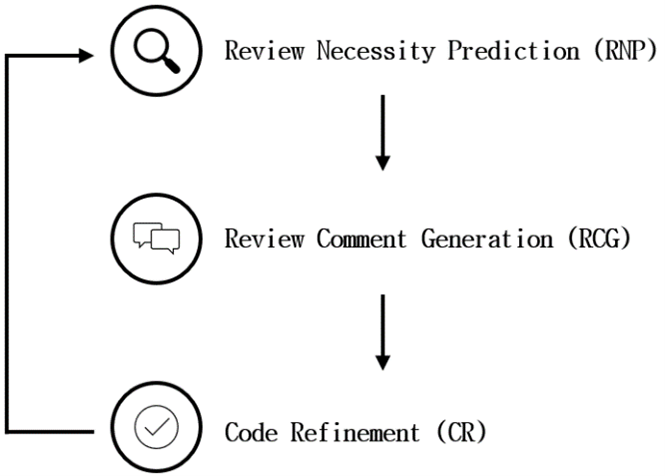}
    \caption{The Cycle of Code Review Process }
    \label{fig:enter-label}
\end{figure}

\subsection{The Performance of GPT on Automated Program Repair}
In the field of Automated Program Repair (APR), GPT Shows results that surpass existing state-of-the-art APR technologies. However, there is a data leakage issues where datasets commonly used in APR tasks are included in the training data, which may lead to overestimation of the performance of LLM. There is a possibility that the performance of LLM may be overestimated due to data leakage when performing bug fixing tasks in a process trained with massive data obtained from the Internet. 

Quanjun Zhang et al. [10] collected competition programming problems and user-submitted codes that did not cause data leaks, built a dataset as follows, and conducted an APR performance evaluation of ChatGPT.

\begin{itemize}
    \item \textbf{Raw data collection}: Crawling exercises and all Java submitted code
    \item \textbf{Obtain fix diffs of buggy code}: keep only pairs with differences of less than 6 tokens between the buggy program and the correct program.
    \item \textbf{Test case extraction}: Download more test cases from the competition’s dedicated database
    \item \textbf{Static-based filtering}: Remove repetitive submissions and comments within code to ensure not to affect GPT judgment
    \item \textbf{Dynamic-based filtering}: Run all remaining submissions and remove pairs if they fail the test cases
\end{itemize}

As a result of the experiment, three results were confirmed in the APR ability evaluation of ChatGPT. First, ChatGPT was able to fix 109 out of 151 bugs when given the default prompt. Second, Chat GPT was able to fix 18, 25, and 10 additional bugs by adding respectively a programming problem description, error message, and bug location to the prompt. Third, ChatGPT was able to fix 9 bugs that were not fixed by default prompts or prompts containing error information.

Based on the above experimental results, it is expected that GPT’s outstanding code modification ability in the APR field can also be utilized in generating code feedback to support programming language learning. Furthermore, it was interesting to discover the impact of the comments in the submitted code on the performance of LLM during the data collection and preprocessing process.

When designing a code review module using GPT, a step to remove comments in code submitted by learners must be included. Additionally, adding a problem description, error message, and bug location to the prompt has proven effective in improving bug fixing performance.

\subsection{LLMs in Computer Science Class}
The recent LLMs, such as the recently released ChatGPT and GitHub Copilot, are receiving significant attention from computer science education experts. Education experts have found that LLMs can generate accurate solutions and accurately explain code for a variety of introductory programming assignments. The outstanding capabilities of these LLMs are generating ongoing discussions about how programming should be taught in computer science classes. At this point, the opinions of computer science instructors are divided. Lau and Guo [11] interviewed 20 introductory programming instructors about how to apply it to their classes. The author responded that in the short term, many instructors will ban the use of LLMs in class to prevent AI-assist cheating. On the other hand, the remaining respondents responded that they would be willing to introduce LLMs into their classes.

Through prior research, LLMs have confirmed the possibility of creating learner-friendly feedback by designing prompts. Therefore, in order to meaningfully utilize LLMs in computer science classes, fit-for-purpose prompt engineering is necessary. In particular, in the introductory programming class, where students learn programming language proficiency and algorithms by solving various programming problems, the rules of using ChatGPT, which includes Codex, a powerful code generation model, are needed. If novice learners use ChatGPT to easily generate code through prompts to solve problems, their learning effectiveness may decrease and they may become dependent on GPT.

To prevent AI-assist cheating side effects that may occur when GPT is introduced into programming language classes, prompts and system flows for generating code review comments need to be designed effectively and systematically. For prompts, Majeed Kazemitabaar [5] suggests including the following principles when designing prompts:
\begin{itemize}
    \item Include at least one input/output example
    \item Structuring the response, including markdown and delimiters for display
    \item Verify the accuracy of submitted code and ensure technical accuracy necessary for code modification
    \item Use a style and tone that does not make students uncomfortable
\end{itemize}

When constructing the code feedback system, it is necessary to effectively display the response results generated by GPT through prompts on the web-front by distinguishing between the function of annotating code lines that require modification based on Markdown and learner-friendly feedback. 

\section{Method}
This study was conducted through the following procedures: dataset collection, prompt template design and enhancement, design guideline establishment, initial system design and development, usability test and improvement, and improved system evaluation. These procedures can be presented in Figure 2. 

\begin{figure} [h]
    \centering
    \includegraphics[width=0.5\linewidth]{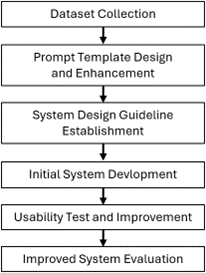}
    \caption{Research Methodology}
    \label{fig:enter-label}
\end{figure}

In the dataset collection stage, test data for testing prompt template was obtained using logs of an existing online judge system. In the prompt template design and enhancement stage, the code review module was designed to develop prompt templates, and then the module was advanced by testing with collected dataset. In the system design guideline establishment stage, improvements were identified and a system design guideline was derived through literature research and comparison with the judge system. In the initial system development stage, the development settings for deploying service was established, the functions of modules were designed and implemented, and UI was built. In the Usability Test and Improvement stage, a usability test was conducted on the initial system and improvement strategy was established based on the results.

Finally, after improving the initial system, the degree of performance of the improved system was evaluated according to research queries, and its suitability as a programming language learning support tool for primary and secondary school students was verified.

\subsection{Dataset Collection}
In order to systematically manage test data, the data frame was composed of eight-labels as follows:
\begin{itemize}
    \item \textbf{Exercise ID (Ex.ID)}: Primary key of Exercise registered in the online judge system
    \item \textbf{Title}: Name of exercise
    \item \textbf{Description (Desc)}: An instruction containing the exercise requirements, input examples, and output examples
    \item \textbf{Solution}: Instructor’s answer code
    \item \textbf{Sub. Code (Submitted Code from students)}: Code submitted by students
    \item \textbf{Solved Subs (The number of Solved Submitted Code)}: The number of correct answers among the codes submitted by students
    \item \textbf{Total Subs (The total number of Submitted Code)}: The total number of codes submitted by students
    \item \textbf{Accuracy}: Proportion rate of correct answers
\end{itemize}

Exercises, students’ submitted codes, and instructors’ answer codes were collected from Company C’s Online Judge System [12] for the introductory Python programming course obtained by referencing the data collection method in the 2023 AtCoder Competition [10] used by Quanjun Zhang et al.

\begin{itemize}
    \item \textbf{Raw data collection}: Starting in 2021 year, among the codes submitted in Python in Company C’s Online Judge System, exercise, student’s submitted code, and the solution were secured
    \item \textbf{Static-based filtering}: Remove repeatedly submitted code and delete comments within the code to ensure they do not affect GPT performance
\end{itemize}

 A total of 93 test data for 27 questions that underwent static-based filtering were entered into the data frame, and an example can be found in Table 1.

\begin{table} [htbp]
\caption{Example of Test Data on Data Frame}
    \includegraphics[width=1\linewidth]{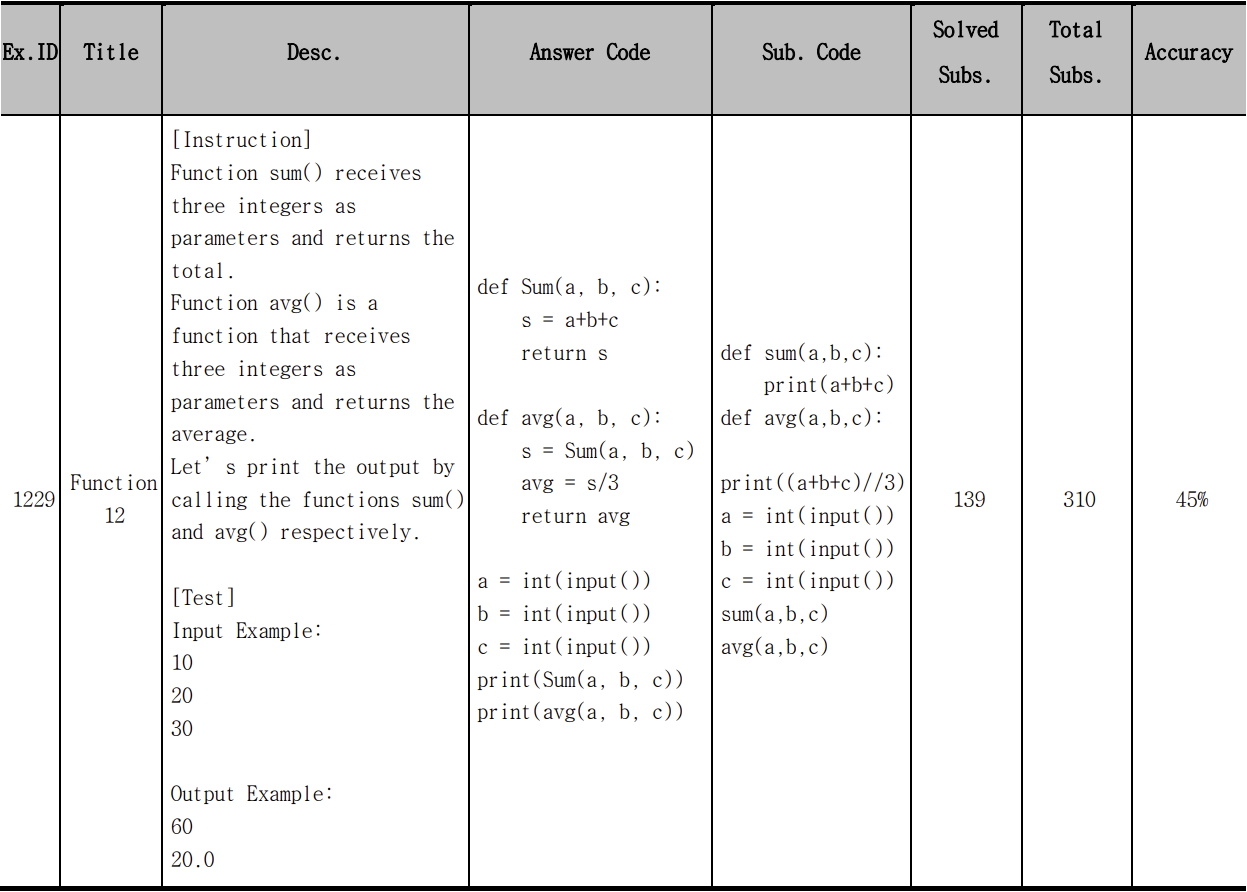}

\end{table}

\subsection{Prompt Template Design and Enhancement}
To Provide code feedback, which is the core function of this system, a code review module was designed, based on the three-stage feedback generation pipeline proposed by J. Lu et al [9]. In order to provide learner-friendly feedback suitable for primary and secondary school students, comments in code reviews are technically accurate. In addition, to be easily distinguished and exposed in the comment display area and code editor area, feedback generated through must be created in markdown format. Furthermore, the style and tone of code review comments must be supportive and constructive, taking into account the target age. The code review module consists of three main prompts (Role setting, Review Necessity Prediction, and Review Comment Generation).
\begin{itemize}
    \item \textbf{Role-setting prompt}: Processes learner’s requests by setting the role of the model
    \item \textbf{Review Necessity Prediction prompt}: Responds with a binary label of “yes” or “no” indicating the need for review of the code submitted by the learner. If “no”, end response
    \item \textbf{Review Comment Generation prompt}: Generates comments for code reviews
    \begin{itemize}
        \item \textbf{Style\&Tone}: Indicates the tone or style of the response
    \end{itemize}
    \begin{itemize}
        \item \textbf{Instruction}: Requirements for adding markdown so that code review results can be distinguished and displayed on the web front
    \end{itemize}
    \begin{itemize}
        \item \textbf{Restriction}: Instruction not to present directly solution code
    \end{itemize}
    \begin{itemize}
        \item \textbf{Exercise}: Description of practice problem selected by the learner in the online judge system
    \end{itemize}
    \begin{itemize}
        \item \textbf{Submitted Code}: Code submitted by the learner
    \end{itemize}
    \begin{itemize}
        \item \textbf{Solution}: Answer code written by the instructor
    \end{itemize}
    \begin{itemize}
        \item \textbf{Example}: Ideal prompt example for RCGP
    \end{itemize}
\end{itemize}

In the process of upgrading the prompt, three main criteria were derived. First, feedback must provide clear and useful value to users. Second, the module structure must be logical and intuitive. Third, modules must be optimized for efficient length while maintaining performance. Based on this standard, the code review model was updated a total of five times, and the following improvements were made during the update process: modifying markdown tags to fit the code review concept, minimizing input-token by changing markdown tags from Korean to English, changing the prompt from conversational type to noun type, improving the structure and order of prompts to ensure a logical and natural flow. 

Through these improvements, the prompt length was shortened from 722 characters including spaces (553 characters excluding spaces) to 701 characters (588 characters excluding spaces) in Korean. In particular, by adding solution and example, which are sub-prompts of RCGP, the hallucination phenomenon of comments was significantly reduced, and it was confirmed that the style and tone of comments were created in a consistent manner. \\
\begin{figure} [h]
    \centering
    \includegraphics[width=1\linewidth]{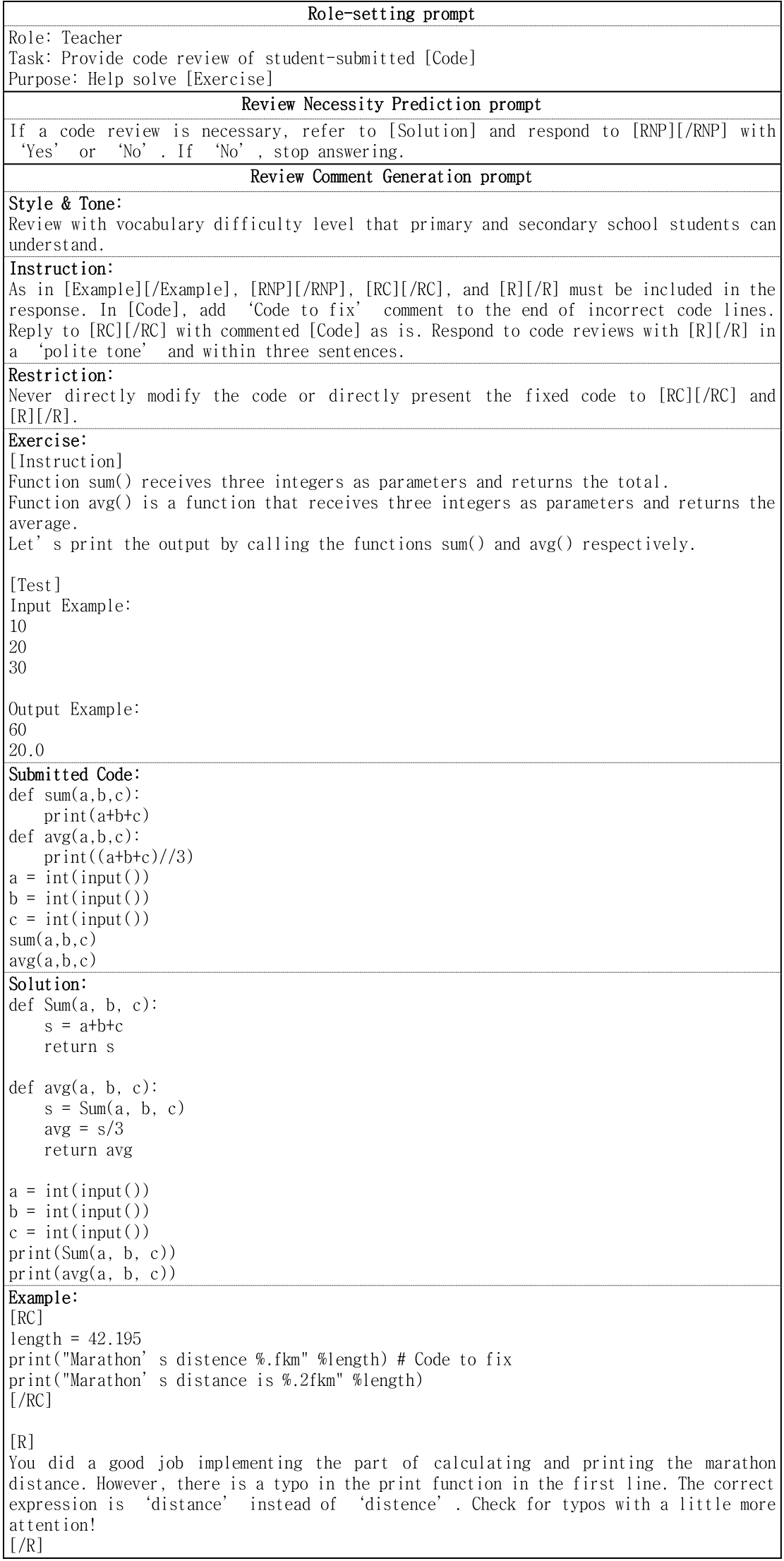}
    \caption{Example of the Code Review Module }
    \label{fig:enter-label}
\end{figure}

\subsection{System Design Guideline Establishment}
Design guidelines were established based on literature research and improvements to the existing judge system. First, learner-friendly feedback must be provided. In the existing judge system, only incorrect answers to problems can be checked, and learners had to identify problems themselves through error messages based on test cases. During this process, learners often experienced difficulties using English and interpreting error messages. By introducing a GPT-4-based AI code review system, it is possible to prevent learning deviation and improve the programming learning experience by providing code review comments that are immediate, accurate, and with a supportive tone.

 Second, comments must be provided on lines that require to fix in the student-submitted code. In the existing judge system, the type of error found during the compilation process and the code line where the error occurred were shown in text, making it difficult for learners in the introductory course to understand the feedback. To solve this inconvenience, when a user clicks the code submission button, a ‘Code to fix’ comment is added to the code line that needs to be fixed through the code review module. This feature allows learners to immediately identify the code where problems occur.

 Third, AI-assist cheating must be limited. The system only provides comments on the code submitted by the learner and does not directly present the fixed code, thereby supporting the learner to think independently and solve problems. Additionally, if the AI-chat is provided directly, learners can easily get the correct answer through the prompt, which may lead to cheating. To prevent this, learners can only receive feedback on their submitted code by clicking the ‘Ask Code Tutor’ button.

 Fourth, an integrated UI must be provided. Presenting problems, writing code, and checking code review comments should be provided on one page to enhance the learning and ease of use.

 Fifth, an editor for Python programming must be provided. The Python style was not applied to the code editor area of the existing judge system, so that reducing the readability of the code. To solve this, an instance of Monaco Editor for Python programming was added. In this editor, learners can program in Python with improved readability and ease of writing.

\subsection{Initial System Development}
\subsubsection{System configuration and environment}
In order to quickly develop this system as a web application and reduce the man-hours of deployment work, the backend and frontend were built integratedly using the Next.js framework. As the system configuration can be seen in Figure 4, a proxy server was set up using Next.js API Routes to solve the CORS (Cross-Origin Resource Sharing) problem that occurs when a client app requests a GPT endpoint in direct.

\begin{figure} [h]
    \centering
    \includegraphics[width=0.75\linewidth]{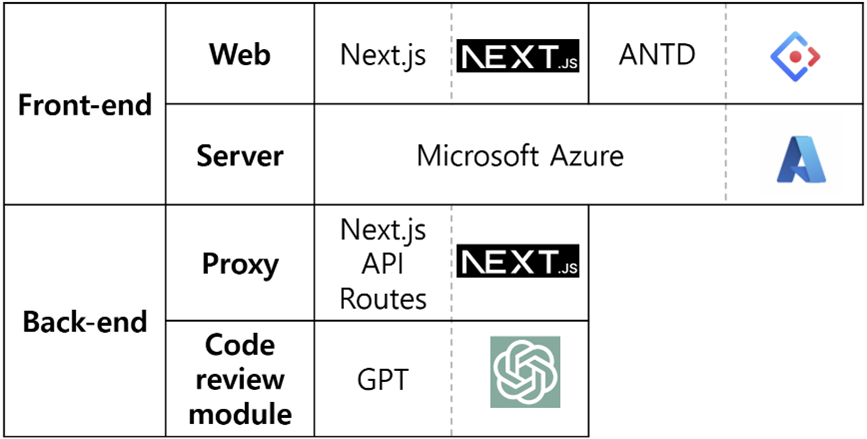}
    \caption{System Configuration and Environment}
    \label{fig:enter-label}
\end{figure}

Backend and frontend include two modules that are responsible for the core functions of this system. First, the Code Correctness Check Module is a tool that checks learners’ submitted code against the correct answer, compiles the code, and provides the result to the learner. Without the need to deploy a Python compiler in the backend, the prompt was configured to enable the GPT-4 to function as the compiler as shown in Figure 5.

\begin{figure} [h]
    \centering
    \includegraphics[width=1\linewidth]{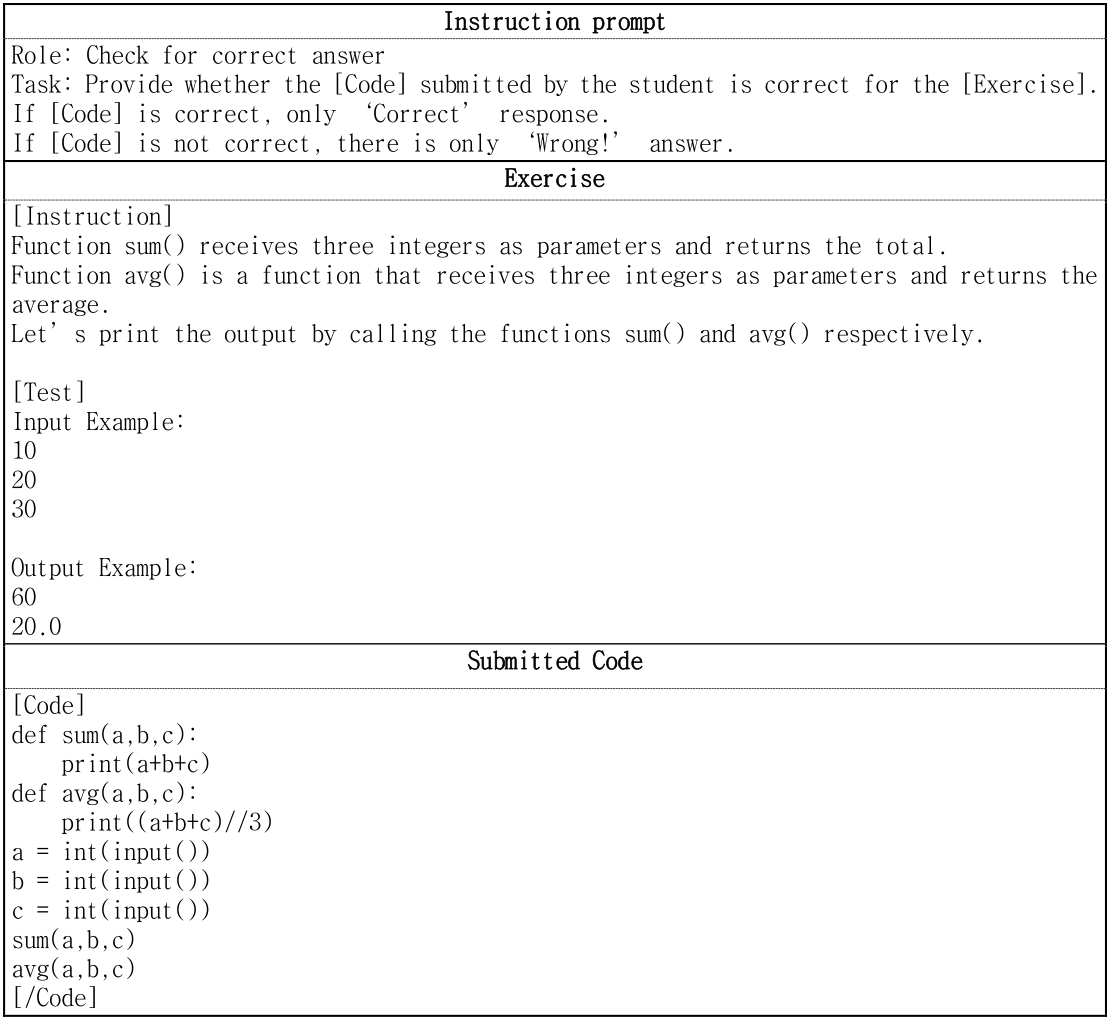}
    \caption{Example of the Code Correctness Check Module}
    \label{fig:enter-label}
\end{figure}

Next, the Code Review Module, which provides code review comments on code submitted by learners, implemented this function by applying advanced prompts to the project five times previously.

The build and deployment of the system were carried out using Microsoft Azure’s Static Web App service. \\

\subsubsection{User Interface and Execution Examples}
Figure 6 shows the UI of the developed system, and the roles and functions of each part are as follows.

\begin{figure} [h]
    \centering
    \includegraphics[width=1\linewidth]{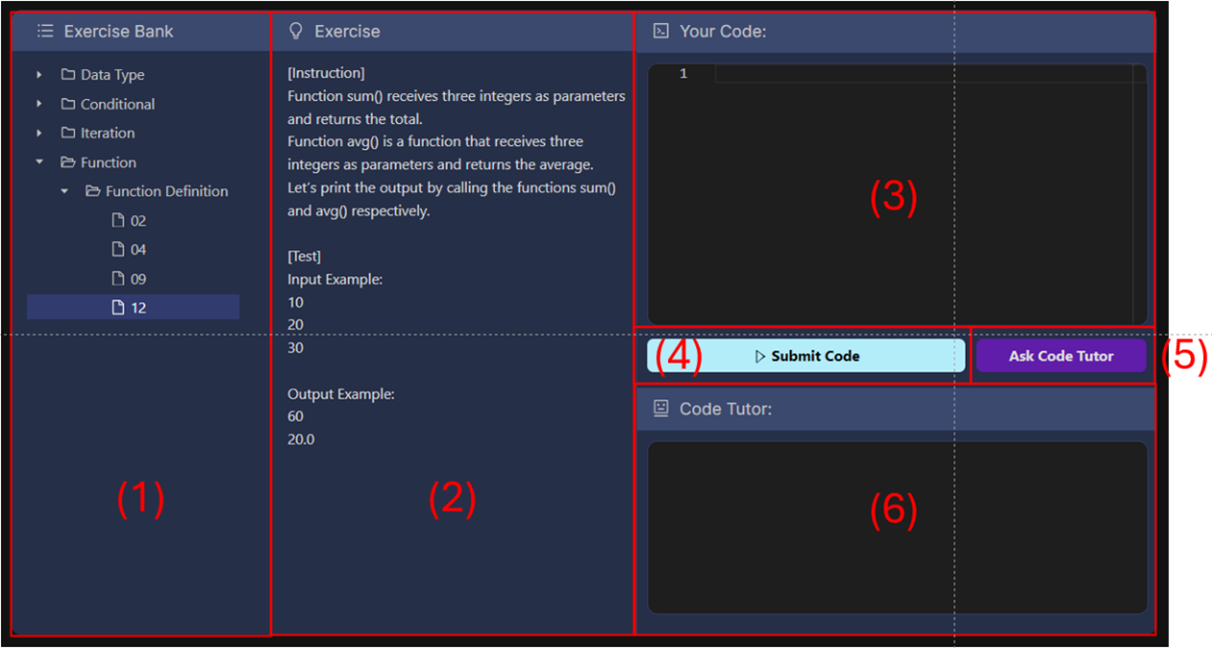}
    \caption{User Interface (UI) }
    \label{fig:enter-label}
\end{figure}

(1) This area provides a exercise choice function organized in a tree format so that learners can examine and select various problems.\\
(2) This area shows the details of the selected problem (the exercise requirements, input examples, and output examples).\\
(3) This is a Python editor for writing code. When a user clicks ‘Ask Code Tutor’, the annotation ‘Code to fix’ is added to the part of the submitted code that needs to be modified, as shown in Figure 7.
\begin{figure} [h]
    \centering
    \includegraphics[width=1\linewidth]{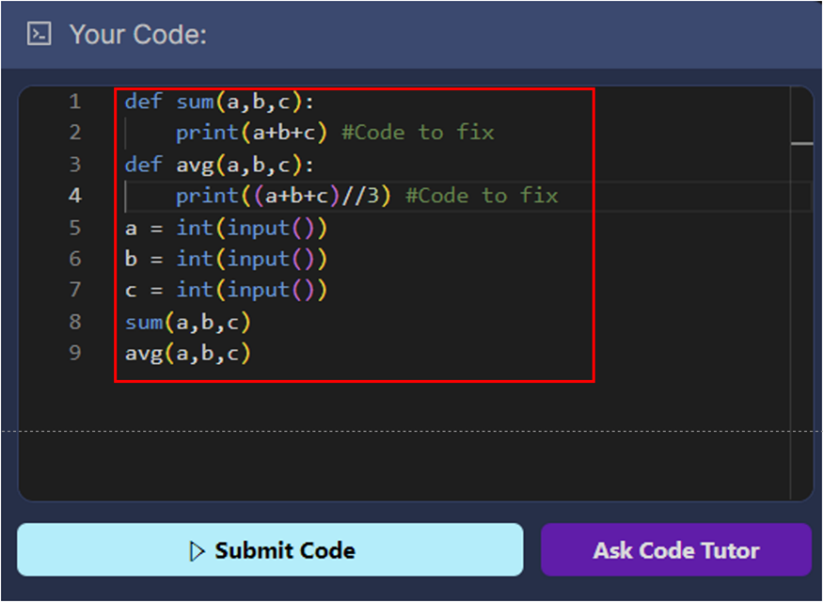}
    \caption{Example of Line Highlighting in Submitted Code }
    \label{fig:enter-label}
\end{figure}
\\
(4) This button submits the code a learner wrote in ‘Your Code’, and the code correctness check module using GPT-4 determines whether the submitted code is correct and displays the correctness and error status in a pop-up as shown in Figure 8. 

\begin{figure} [h]
    \centering
    \includegraphics[width=0.9\linewidth]{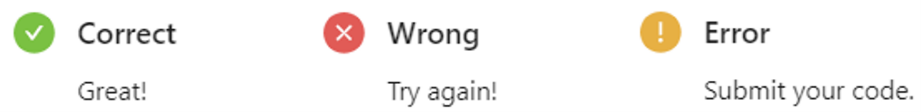}
    \caption{Example of the Three States for Answer Validation}
    \label{fig:enter-label}
\end{figure}
(5) This button requests code review comments through the GPT-4 endpoint using the Code Review Module.\\

(6) This area shows the feedback received from GPT-4, which can be seen in Figure 9. 

\begin{figure} [h]
    \centering
    \includegraphics[width=1\linewidth]{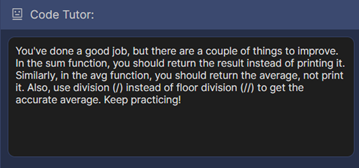}
    \caption{Example of Code Review Comments Displayed  }
    \label{fig:enter-label}
\end{figure}

\subsection{Usability Test and Improvement}
In order to verify the initial system, a usability test was conducted on this system, which is scheduled to be used for primary and secondary school students. Three software education experts participated in this evaluation, all of whom had at least two years of programming education experience. The evaluation was conducted by collecting qualitative feedback through interviews conducted after participants fully used the system. The evaluation items were interview questions based on web based artificial intelligence application service elements proposed by Moon Hee-Jeoung [13], including functions and services, user experience, troubleshooting and support, automated process, and sentiment and opinion analysis. 

\subsubsection{Analyzing Usability Test Feedback}
Based on the results of the usability test feedback, the system improvement strategy was summarized, and priorities for advancement were selected focusing on issues related to the code tutor. The contents are summarized in Table 2. 

\begin{table} [htbp]
\caption{Analyzing Feedback and Improvement Strategies}
    \includegraphics[width=1\linewidth]{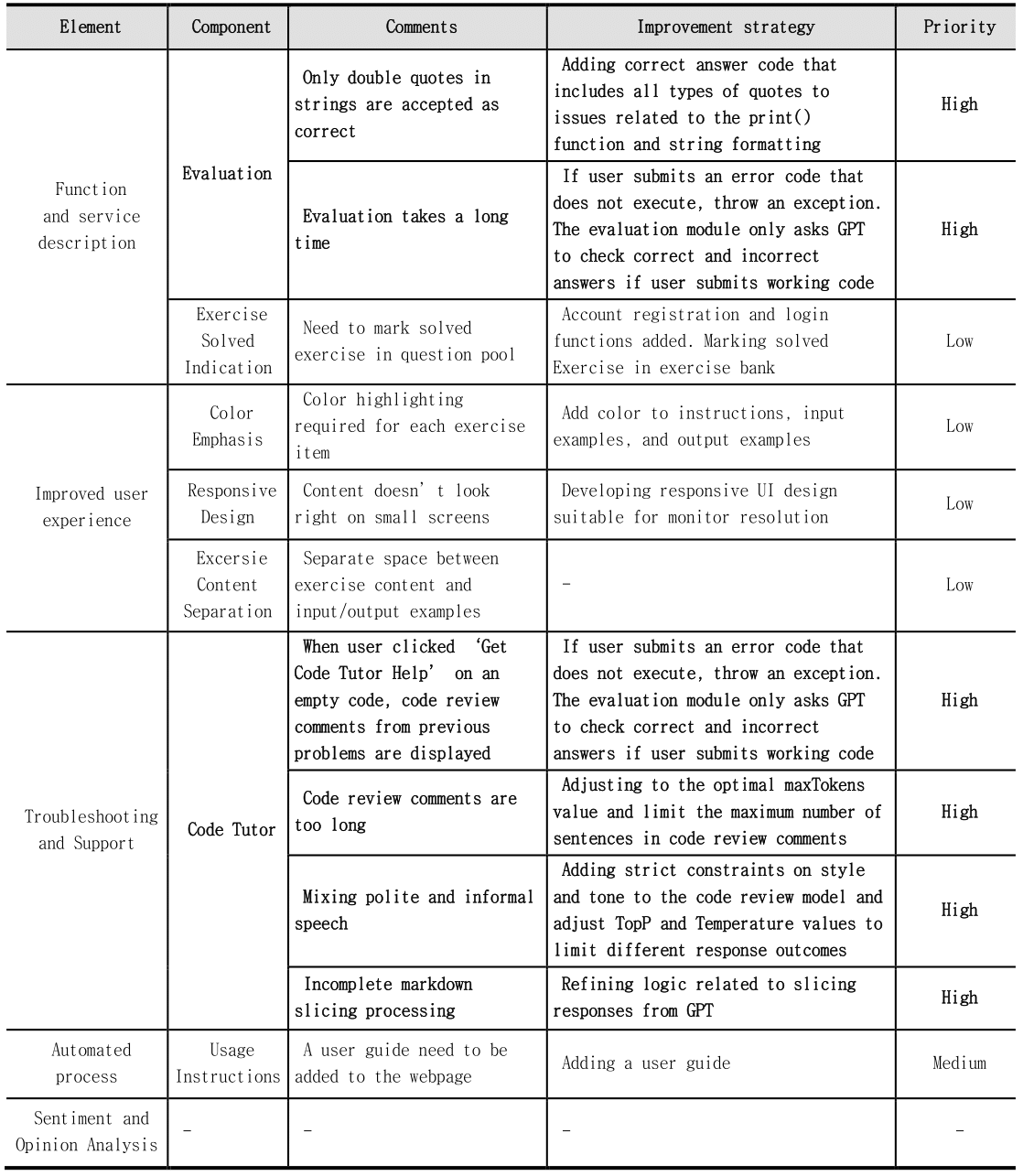}

\end{table}

Regarding functions and services, issues were identified in which only double quotation marks were recognized as correct in strings and that grading time was prolonged. To improve this, the code correctness check module is advanced to allow various correct answers to submitted code, and the system flow is advanced to shorten the response time of the module.

Concerning user experience, there was a problem with poor readability of problem areas and content not being visible properly on small screens. To improve this, colors are added to questions, input examples, and output examples in the problem area to improve readability, and a responsive UI design that matches the resolution of desktop and laptop monitors is needed.

In troubleshooting and support, there were issues with code review comments from previous issues appearing when clicking ‘Ask Code Tutor’ in an empty code state, the response result being too long, making it difficult to understand the content, and issues with formal and informal language being used interchangeably. Accordingly, exception handling is added to prevent code review comments from being generated in empty code states, and the module is improved to provide code review comments concisely. And add a constraint to respond politely to the code review comment module.

In the automated process, users experienced inconvenience because they did not know how to use the system before using it. To solve this problem, it is necessary to add a how-to guide and present how to use it through videos.
\\
\subsubsection{Enhancing the Initial System for Improved Usability}
Items previously marked as high priority related to code tutor and code correctness check have been improved. The flow chart of the improved system can be seen in Figure 10.

\begin{figure} [h]
    \centering
    \includegraphics[width=1\linewidth]{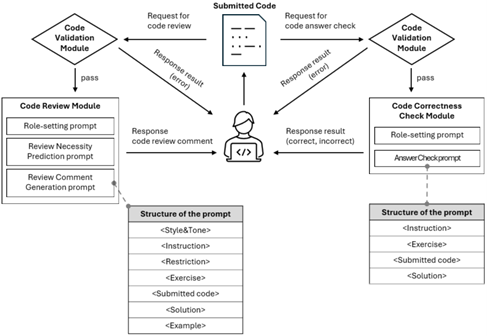}
    \caption{System Flow of Improved Version}
    \label{fig:enter-label}
\end{figure}

First, the improvement in the code correctness check process is that when a learner writes code and submits the code, the newly added ‘Code Validation Module’ checks whether it is properly functioning Python code. If there is a problem with the code, the user is notified that the code is incorrect and a code correctness check is not performed. On the other hand, if the code validation module does not find any problems with the code submitted by the learner, it makes a request to the GPT endpoint through the code correctness module.

Next, the improvement in the process of requesting code review comments is that when a learner writes code and submits the code, the code validation module checks whether it is a properly functioning Python code. If there is a problem with the submitted code, it is shown to the learner that the code is incorrect. Otherwise, feedback is requested from the GPT endpoint through the code review comment module.

At this time, the maxTokens value was optimized to reduce response delay, and both the number of tokens in responses including markdown and the maximum number of sentences in code review comments were limited. The values of the temperature and topP parameters were adjusted to reduce response time and generate consistent style and tone.\\\\

\section{Improved System Evaluation}
This section verifies how well the improvement strategy derived from the usability test was applied to the improved system through answers to the four research queries presented above. A performance evaluation was conducted at three times in order: strict code correctness check in improved system, response time reduced in code reviews, and cost reduction via API call optimization.  Lastly, software education experts participate in a survey to comprehensively evaluate the quality of code reviews. 

Specifically, \textbf{\textit{RQ1}}: Are the results of the GPT-based Code Correctness Check more strict than existing online judge system?, \textbf{\textit{RQ2}}: Does the improved system reduce the response time for code review comments compared to the existing system?, \textbf{\textit{RQ3}}: Does the improved system reduce API call costs compared to existing system?, \textbf{\textit{RQ4}}: Despite reduced response times and call costs, does the system maintain the quality of code reviews?\\

\subsection{RQ1: Strict Code Correctness Check in Improved System}
After analyzing system logs from the online judge system, four main types of errors in code submitted by learners is identified:
\begin{itemize}
    \item Unnecessary Code: Code that is not essential for solving the problem has been added
    \item Requirement not met: When the submitted code ignores a specified requirement and is solved differently
    \item Hard Coding: When the result is coded hard using the input/output example as is without proper logic
    \item Computation Error: Calculation errors caused by complex logic
\end{itemize}

A total of 108 test codes, which were successfully reviewed by the improved system for 27 issues containing these error types, were collected. Additionally, the failure rate was measured for each error type. The failure rate was calculated by dividing the number of test codes that passed in the improved system (N) by the number of test codes that failed for that error type (ni) to calculate the failure rate (Ri) for that error type.\\
\begin{equation}
Ri = \left(\frac{ni}{N}\right) \times 100
\end{equation}
\\\\As can be seen in Table 3 and Figure 8, the improved system was able to identify errors more effectively than the existing system. In particular, the detection rate of hard coding (21.3\%) and unnecessary code (17.59\%) was particularly high.

\begin{table} [htbp]
\caption{Error Type Failure Rates in the Online Judge System }
    \includegraphics[width=1\linewidth]{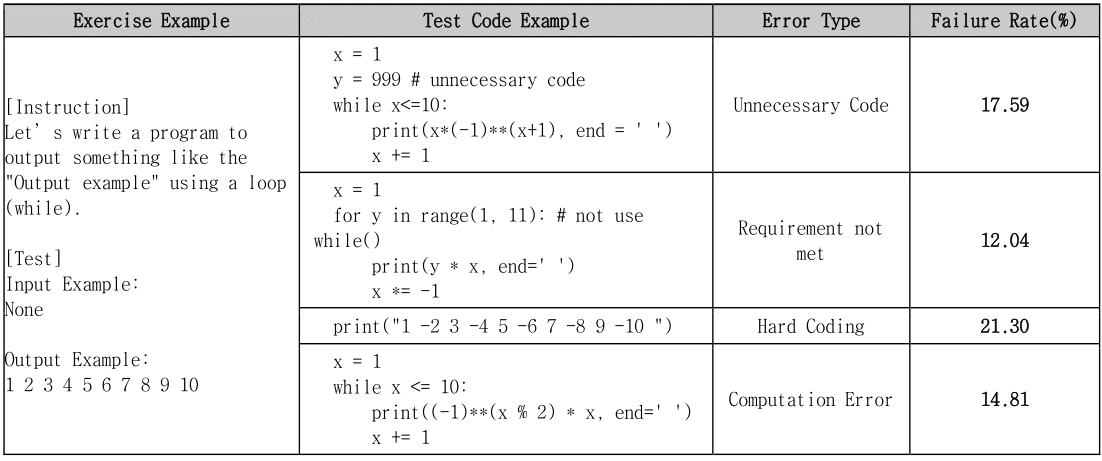}

\end{table}

\begin{figure} [h]
    \centering
    \includegraphics[width=1\linewidth]{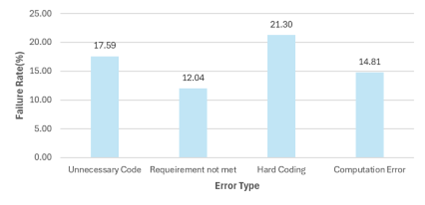}
    \caption{Error Type Failure Rates in the Online Judge System}
    \label{fig:enter-label}
\end{figure}

\subsection{RQ2: Response Time Reduced in Code Reviews}
In this study, the response time was measured by running 92 error codes submitted for 27 questions of the existing system in the code correctness check module of the initial system and improved system. As can be seen in Table 4 and Figure 12, the improved system was able to reduce the response time by at least 12\% and up to 58\% compared to the initial system.

\begin{table} [htbp]
\caption{Response Time between Initial and Improved System}
    \includegraphics[width=1\linewidth]{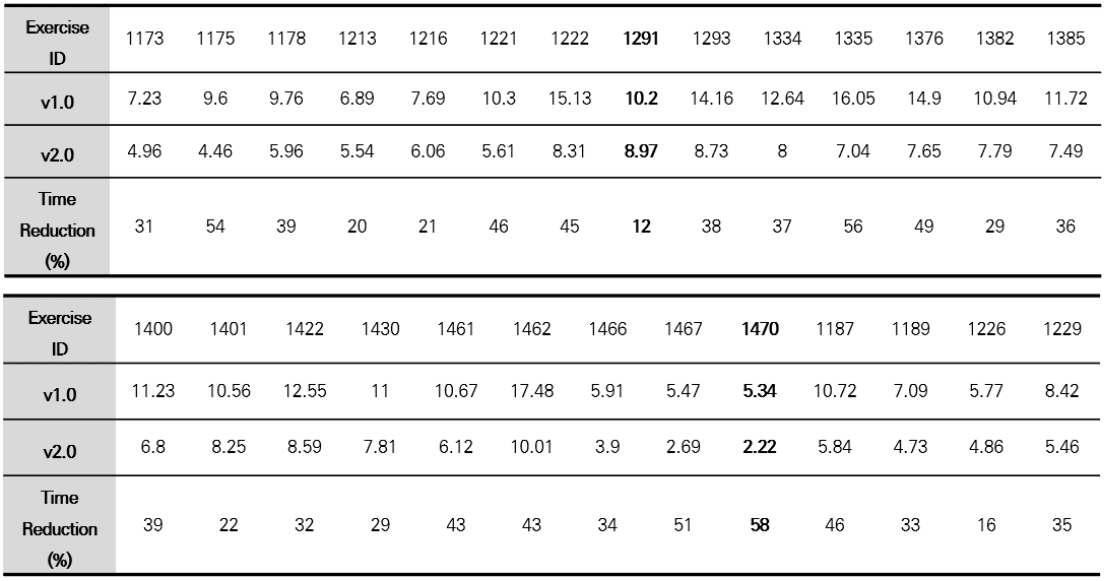}

\end{table}

\begin{figure} [h]
    \centering
    \includegraphics[width=1\linewidth]{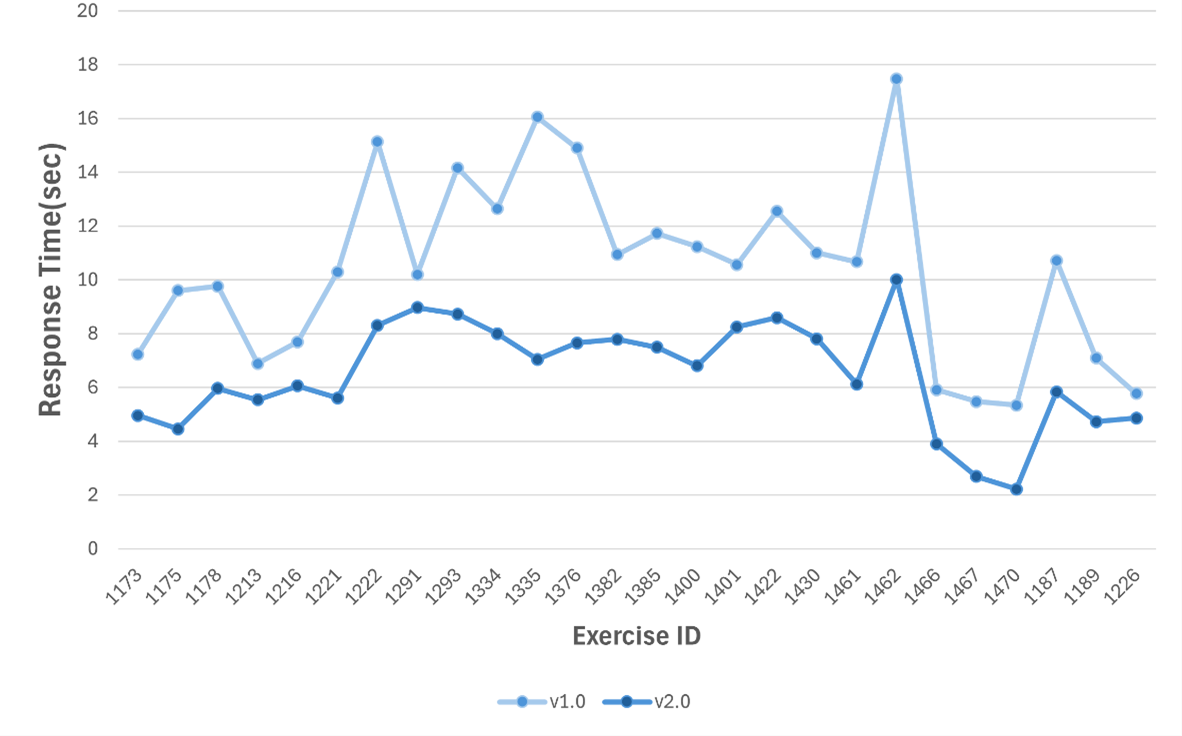}
    \caption{Response Time for Code Review Comment Requests}
    \label{fig:enter-label}
\end{figure}

\subsection{RQ3: API Call Cost Reduction via Optimization}
In this study, test data from all 27 questions were used to measure input tokens and output tokens occurring in the code review comment module of the initial system and improved system. After calculating the average value of tokens in the two systems, the API call cost was estimated in US dollars (USD) based on OpenAI’s pricing[14]. As can be seen in Table 5 and Figure 13, the improved system was able to reduce the OpenAI API call cost by up to 8.53\% compared to the initial system.

\begin{table} [htbp]
\caption{Cost Per Call between Initial and Improved System}
    \includegraphics[width=1\linewidth]{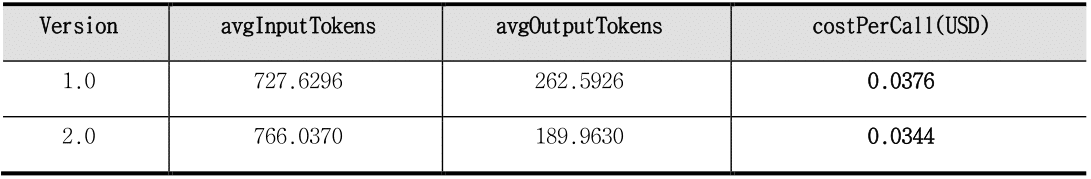}

\end{table}

\begin{figure} [h]
    \centering
    \includegraphics[width=1\linewidth]{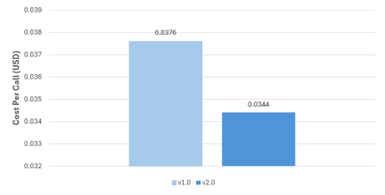}
    \caption{Cost Per Call for Code Review Comment Requests}
    \label{fig:enter-label}
\end{figure}

\subsection{RQ4: Quality of Code Reviews}
A survey was conducted to verify the improved system’s code review comments and responses. This was done by referring to the verification and feedback examples suggested by Choi Seong-yune et al [4]. From this, five evaluation criteria were derived: precision, usefulness, specificity, supportive tone, and learning effect.

Precision refers to whether the code tutor accurately points out errors in the code and suggests solutions, and whether lines of code that need correction are accurately annotated in the Your Code area. Usefulness is verification that the code tutor’s comments actually help solve the problem. Specificity is ensuring that the code tutor’s comments support specific errors or problems. The supportive tone of the feedback ensures that the code tutor’s comments are not overly critical. Lastly, learning effect is a standard that checks whether learners can understand new concepts or improve their code writing skills through the code tutor’s comments.

Using the 5-point Likert scale, a total of 6 people, including 3 current programming education instructors who previously participated in the system evaluation, 2 additional programming education instructors, and 1 coding content planning developer, conducted the evaluation through a survey. The evaluation was verified by submitting correct and incorrect answer codes at least twice for each exercise. As can be seen from the survey results in Table 6 and Figure 14, the overall response to the code review comments was positive, but opinions varied in some specific areas. Additionally, participants consistently evaluated this tool as suitable for programming learning for elementary and secondary school students.

\begin{table} [htbp]
\caption{Survey Results of Code Review Comments Quality}
    \includegraphics[width=1\linewidth]{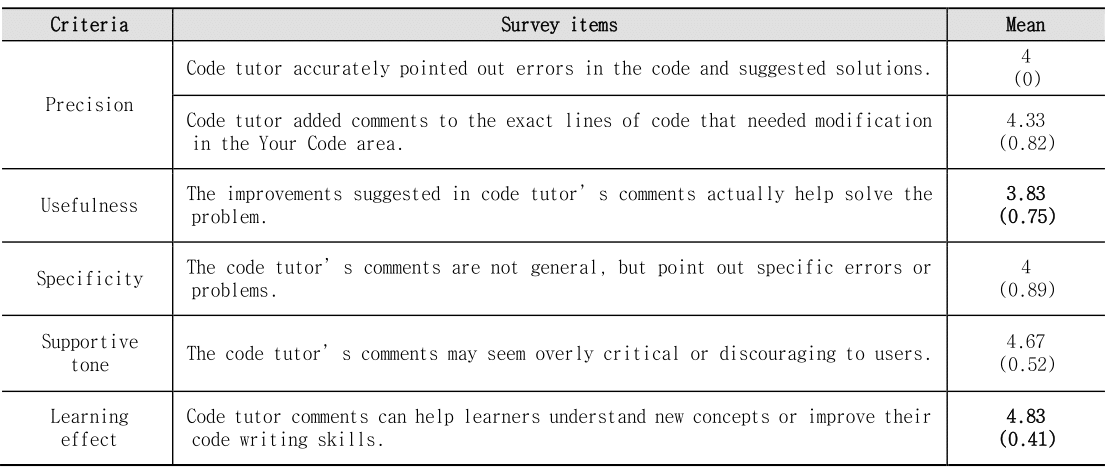}

\end{table}

\begin{figure} [h]
    \centering
    \includegraphics[width=1\linewidth]{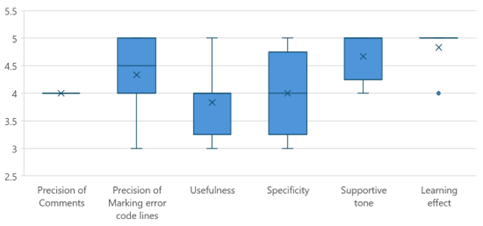}
    \caption{Distribution of Survey Results of Code Review}
    \label{fig:enter-label}
\end{figure}

\section{Conclusion and Further Study}
The improved system is expected to be useful as a programming language  learning tool for primary and secondary school students. Compared to the existing online judge system, this system was found to provide results by strictly checking error types such as unnecessary code, requirement not met, hard coding, and computation error. The improved system was confirmed to shorten response times and API call costs through prompt engineering of the code review comments module and newly added code validation modules, and by optimizing the length of output tokens. Additionally, although response time and call costs were reduced, there were no major problems with the quality of service of code review comments.

In follow-up research, it is necessary to further improve usability, introduce a membership management system, and verify the effectiveness of the system with primary and secondary school students. The improved system can be found at https://www.codetutor119judge.com/.\\

\end{document}